\documentstyle[sprocl]{article}

\def\be{\begin{equation}}
\def\ee{\end{equation}}
\def\bea{\begin{eqnarray}}
\def\eea{\end{eqnarray}}

\begin{document}

\title{Implications of Temporal Structure in GRBs }
\author{Tsvi Piran and Re'em Sari}
\maketitle

\address{Racah Institute of Physics, The Hebrew University,\\ 
Jerusalem,  91904 ISRAEL} 
\abstracts{\ The highly variable temporal structure observed in most GRBs
provides us with unexpected clues. We show that variable GRBs cannot be
produced by {\it external shocks }models and consequently cannot be produced
by an ``explosive'' {\it inner engine. }The observed temporal structure must
reflect the activity of the {\it inner engine }that must be producing
unsteady and irregular ``wind'' which is converted to radiation via {\it %
internal shocks.} }

Cosmological and even distant galactic halo \cite{PiS93} GRBs models must
overcome the compactness problem. Relativistic effects provide the only
known solution to this problem. According to the generic picture \cite
{Piran97} GRBs arise from a three stages processes: (i) A compact {\it inner
engine }produces a relativistic energy flow (relativistic particles or
electromagnetic Poynting flux - but not the observed photons) which (ii)
transport the energy outwards to an optically thin region where (iii) it is
converted to the observed radiation. Step (iii) could occur due to {\it %
external shocks} \cite{MR1}- resulting from the deceleration of the energy
flow into some external medium (e.g. the ISM). Alternatively, {\it internal
shocks} \cite{NPP,MR4} that would arise in an irregular flow with non uniform
velocities could convert the kinetic energy to radiation. We show here that 
{\it internal shocks }rather than {\it external shocks }convert the energy
in GRBs.

Extensive efforts have been devoted to the question of how the observed
spectrum is produced, probably because a single spectrum seems to be a
universal characteristic of GRBs. The temporal structure, which varies
drastically from one burst to another, was practically ignored until recently %
\cite{SaP,SNP,Fen,SaP97}. Most bursts have a highly variable temporal
profile with a rapid variability, on a time scale $\delta T\ll T$, $T$ being
the burst's duration. We suggest here that this temporal structure may
provide the clue to this mystery \cite{SaP97}.

The {\it inner engine} produces the energy flow which cannot be observed
directly. The observed $\gamma $-rays emerge only from the outer energy
conversion regions. This poses an additional difficulty in deciphering the
origin of GRBs. Different engines could produce GRBs provided that they can
produce the required relativistic energy flux. Variability on a time scale $%
\delta T$ in the bursts dictates an upper limit to the size of this {\it %
inner engine} $\sim c\delta T$. {\it Inner engines} can be ``explosive'' \cite
{Goodman86} producing a single outgoing shell whose width $\Delta $ is
comparable to the size of the {\it inner engine}. An {\it inner engine} can
also produce a ``wind'' \cite{Pac86}: an outgoing flows on scales longer than
the size of the source. We show here that the observed temporal structure
cannot be produced within the energy conversion regions and it must reflect
the activity of the source. Consequently ``winds'' rather than
``explosions'' power GRBs.

Consider a relativistic shell that converts its energy to radiation. Let $%
\Delta $ be the width of the shell and let the conversion take place between 
$R_{E}$ and $2R_{E}$. The emitting material moves with a Lorenz factor %
\footnote{%
Note that $\gamma $ could be smaller than the initial Lorentz factor of the
shell.}, $\gamma $. There are three generic time scales. First is the
difference in arrival time between two photons emitted at $R_{E}$ and $%
2R_{E} $, $T_{R}\approx R_{E}/\gamma ^{2}c$. Angular spreading, that is
blending of emission from regions from an angle $\theta $ from the line of
sight leads to a second time scale $T_{\theta }\approx R_{E}\theta ^{2}/c$.
Finally, $\Delta /c$ the light crossing time of the relativistic shell,
corresponds to the time difference between the photons emitted from the
shell's front and from its back.

Examine now a very thin shell. A typical source that produces a thin shell
is an ``explosive'' fireball for which the width of the shell is comparable
to the size of the {\it inner engine.} However, sources that produces a
short wind are also of this kind. Now more specifically require that the
shell satisfies: $\Delta \le R_{E}/\gamma ^{2}$. It is remarkable that even
arbitrarily thick shells will satisfy this conditions if the emission is due
to {\it external shocks} \footnote{%
Strictly speaking this was shown only for hydrodymanic shocks \cite{SaP97}.}.
Since $\Delta <R_{E}/\gamma ^{2}$ the duration of the burst is determined by
the energy conversion region and not by the duration that the {\it inner
engine} operates (which determines $\Delta $).

Because of relativistic beaming an observer detects radiation from an
angular scale $\gamma ^{-1}$ around the line of sight. Thus the angular size
of the observed regions always satisfies $\theta \le \gamma ^{-1}$. If the
system is ``spherical'' (that is spherical on a scale larger than $\gamma
^{-1}$) $\theta \approx \gamma ^{-1}$ and then $T_{\theta}\approx
R_{E}/\gamma ^{2}c\approx T_{R}\approx T$. Thus angular spreading will erase
all temporal structure on scales shorter than $T_{\theta}$ resulting in $%
\delta T\approx T$. In order to produce variable bursts with $\delta T\ll T$%
, within the external shock scenario, one must break the spherical symmetry
on scales smaller than $\gamma ^{-1}$.

It is useful to define a variability parameter ${\cal V}\equiv T/\delta
T\sim 100$. Detailed analysis \cite{SaP97} shows that the emitting regions
must have an angular size smaller than $(\gamma {\cal V})^{-1}\le 10^{-4}$
to produce such a variability. A sufficiently narrow jet can bypass this
restriction. But it is not clear how one can accelerate and collimate it.
Hydrodynamic acceleration, for example, cannot produce an angular width
smaller than $\gamma ^{-1}$. A second possibility is an emitting region made
of numerous small size bubbles. The number of bubbles (emitting regions)
should be smaller than ${\cal V}$, otherwise the contribution from different
bubbles will average out to a smooth signal. The maximal solid angel of each
bubble is $(\gamma {\cal V})^{-2}$. Therefor the total solid angel of all
bubbles is smaller than $(\gamma ^{2}{\cal V})^{-1}$, which is only ${\cal V}%
^{-1}$ of the observed solid angle. This leads to an intrinsic inefficiency
in conversion of energy to radiation of magnitude ${\cal V}^{-1}$, ruling
out models based on this idea.

Let's turn now to a wide relativistic shell in the form of a wind. If the
wind is irregular the energy conversion would be due to {\it internal shocks}
and the condition $T_{R}<\Delta /c$ would be satisfied. This will produce a
burst whose overall duration is $\Delta /c$ and the observed variability
scale is \footnote{%
This is provided, of course, that the cooling time is shorter than $%
T_{\theta}$ \cite{SaP97a}.} $\delta T\approx T_{\theta }\approx T_{R}$. The
variability scale could be much shorter than the duration. The duration is
determined now by the activity of the {\it inner engine }and not by the
emitting regions. The observed temporal structure reflects the activity of
the inner engine, which must be producing a relatively long and highly
irregular wind.

We find that only this second possibility, of a ``wind'' like {\it inner
engine }and energy conversion by {\it internal shocks}, can produce the
observed temporal structure. This conclusions have several direct
implications. First it tells us that the emitting regions operate with the 
{\it internal shock} mechanism \footnote{%
In fact we have shown here that {\it external shocks }cannot produce the
observed temporal structure. We have not shown yet that {\it internal shocks 
}can produce it. This work is in progress now.}. This would have direct
implications for any attempts to calculated the observed spectrum from these
events. The implications for the {\it inner engine }are even more dramatic:
It must operate for a long duration, up to hundred of seconds in some cases,
and it must produce highly variable winds as required to form {\it internal
shocks }and the observed variable activity. This directly rules out all
explosive models. We will discuss elsewhere the implication of this
conclusion for some specific models.

We thank Ramesh Narayan and Jonathan Katz for helpful discussions. The
research was supported by a US-ISRAEL BSF grant and by NASA grant NAG5-3516. 

\end{document}